\documentclass[prb,twocolumn,showpacs,amsmath,superscriptaddress,amssymb]{revtex4}

\usepackage{graphicx}
\usepackage{dcolumn}
\usepackage{bm}

\begin{document}


\title{Completely flat bands and fully localized states on surfaces of  anisotropic diamond-lattice 
models}

\author{Ryuji Takahashi}
 \affiliation{Department of Physics, Tokyo Institute of Technology,
2-12-1 Ookayama, Meguro-ku, Tokyo 152-8551, Japan}
 \affiliation{Department of Applied Physics, University of Tokyo,
7-3-1 Hongo, Bunkyo-ku, Tokyo 113-8656, Japan}

\author{Shuichi Murakami}
\affiliation{Department of Physics, Tokyo Institute of Technology, 
2-12-1 Ookayama, Meguro-ku, Tokyo 152-8551, Japan} 
\affiliation{TIES, Tokyo Institute of Technology, 
2-12-1 Ookayama, Meguro-ku, Tokyo 152-8551, Japan} 

\date{\today}

\begin{abstract}
We discuss flat-band surface states on the (111) surface in the tight-binding model with nearest-neighbor hopping on the diamond lattice, in analogy to the flat-band edge states in graphene with a zigzag edge. The bulk band is gapless, and the gap closes along a loop in the Brillouin zone. The verge of the flat-band surface states is identical with this gap-closing loop projected onto the surface plane. When anisotropies in the hopping integrals increase, the bulk gap-closing points move and the distribution of the flat-band states expands in the Brillouin zone. Then when the anisotropy is sufficiently large, the surface flat bands cover the whole Brillouin zone. Because of the completely flat bands, we can construct surface-state wavefunctions which are localized in all the three directions.
\end{abstract}

\maketitle 
\section{introduction}
Flat bands have been studied particularly in the context of possible ferromagnetism driven by interactions, as was proposed by Lieb \cite{Lieb89}, and successively by Mielke and Tasaki \cite{Mielke91a,Mielke91b,Mielke91c, Tasaki93}.
On the other hand, from the research on graphene \cite{Novoselov04} it is known that the tight-binding model with nearest-neighbor 
hopping on a honeycomb lattice with a zigzag edge exhibits flat-band edge states \cite{Fujita96}, 
and its origin is topologically interpreted \cite{Ryu02}.
In the dispersion of a graphene ribbon with zigzag edges, the flat-band edge states appear between 
the wavenumbers corresponding to the projection of Dirac points at K and K'.
In contrast, there are no flat-band edge states in the graphene ribbon with armchair edges, 
because
in the projection of the dispersion, Dirac cones at the K and K' points 
overlap each other. 
In three dimensions, the nearest-neighbor tight-binding model on the diamond lattice, i.e. a three-dimensional analogue
of the honeycomb lattice, exhibits flat-band surface states \ \cite{Takagi08}.

When the hopping of the tight-binding model on the honeycomb lattice becomes anisotropic,
 the Dirac points in the bulk Brillouin zone (BZ) move away from
the K and the K' points. Moreover, when the anisotropy is sufficiently large, the two Dirac points meet and the bulk dispersion relation becomes linear in one direction and quadratic in the other \cite{Dietl08}. In that case, the flat-band edge states cover the whole one-dimensional (1D) BZ \cite{Delplace11}. With a further increase of the anisotropy, the bulk becomes gapped while the completely flat band remains in the edge BZ.

In this paper, we focus on surface flat bands in the nearest-neighbor tight-binding models on the diamond lattice with the (111) surface.
When the model has no anisotropy, the gap closes along 
a loop in the bulk BZ. If the Fermi energy is set to be zero, which corresponds to the case 
with particle-hole symmetry, the bulk Fermi surface (FS) coincides with this loop. If we introduce 
anisotropy in the nearest-neighbor hopping integrals, we find that the loop is deformed and 
shrinks. 
Similarly to the honeycomb-lattice model, the surface flat bands are formed in the ${\bf k}$
region surrounded by the projection of the FS loop. 
When the anisotropy is sufficiently large, the FS loop vanishes, and consequently the flat-band surface states cover the whole 2D BZ.

In both the honeycomb-lattice and the diamond-lattice models, because the edge/surface band is completely flat over the entire BZ, any linear combinations of the edge/surface 
states remain eigenstates. Thereby we can construct edge/surface states that are spatially localized in all directions, i.e. not only along the 
direction into the interior, but also along the edge/surface.
We call such states as fully localized states. 
In addition, we also find that the isotropic case is at a 
topological transition of the bulk FS; the loop of the bulk FS changes its topology by varying anisotropy of the hopping integrals.

The organization of the paper is as follows. In Sec.~II we review how the 
flat-band edge states of the tight-binding model on the honeycomb lattice evolve
with changes in the anisotropy. We discuss analogous behaviors of surface states
of the model on the diamond lattice in Sec.~III. In Sec.~IV, we show how the behaviors of the edge/surface states shown so far are explained by topological argument. Section V is devoted to a
calculation of the edge/surface states that are fully localized, for both the honeycomb- and diamond-lattice models. We summarize our results in Sec.~VI.

\section{Honeycomb lattice}
\begin{figure}[htbp]
 \begin{center}
 \includegraphics[width=80mm]{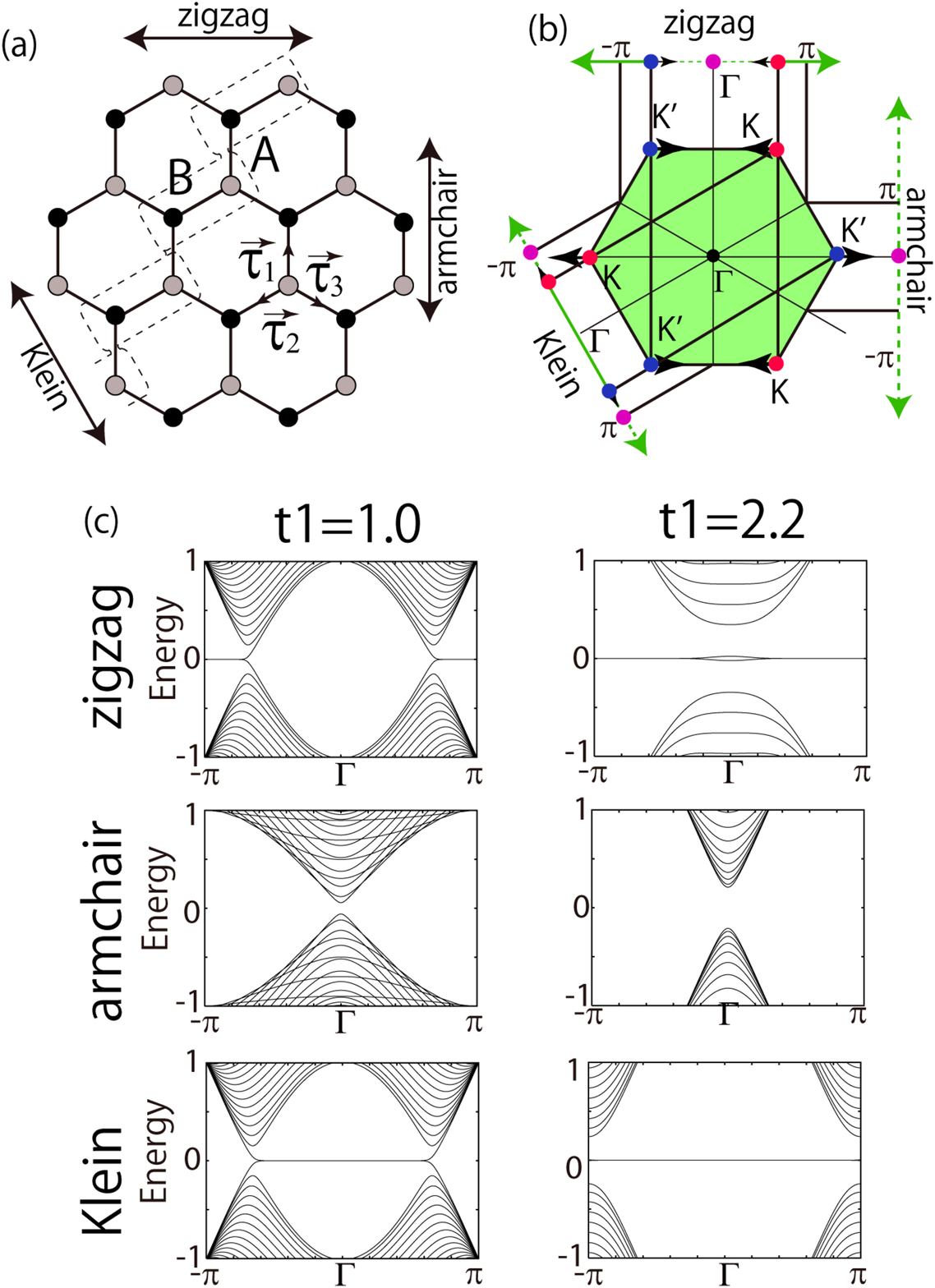}
 \caption{(a) Schematic of the honeycomb-lattice structure. The dotted line shows a choice of unit cell with
translational symmetry along the edge.
The arrows show the directions along zigzag, armchair, and Klein edges.
(b) shows the first BZ of the honeycomb lattice. Dots at the zone corners show the gap-closing points for the graphene model ($t_1=t_2=t_3=1$). 
Namely, the bulk bands become gapless at K and K' points, when the hopping integral is isotropic.
 By increasing $t_{1}$ from unity, the gap-closing points move away from K and K' points as shown by the arrows. The flat-band edge states expand in the BZ as
the bulk gap-closing points move. 
(c) shows the dispersions for ribbon geometry for $t_1=1$ and $2.2$ at $t_2=t_3=1$.
The dispersions are shown for zigzag, armchair and Klein edges.
In zigzag and Klein edges, flat bands appear at the zero energy. }
\label{fig:graphene-like}
\end{center}\end{figure}

We first review the flat-band edge states on 
the honeycomb-lattice structure shown in Fig.~\ref{fig:graphene-like}(a), 
and study the completely flat band for the models with anisotropy, which has been studied in Ref.~\onlinecite{Delplace11}.
We consider a tight-binding Hamiltonian on this lattice,
\begin{eqnarray}
H_{\rm h}=\sum_{\langle ij \rangle}c^{\dagger}_{i}t_{ij}c_{j},
\label{eq:graphene}
\end{eqnarray}
where the subscript ``h'' represents the honeycomb lattice, $t_{ij}$ is the hopping integral along the nearest-neighbor bond vector $\boldsymbol{\tau}_{a}$, and $c_i$ ($c^{\dagger}_{i}$) is the annihilation (creation) operator of the electron.
We treat the hopping integral $t_{ij}$ as a real positive parameter, and it is labeled with the vectors $\boldsymbol{\tau}_{a}$ as $t_{a}$.
The bulk Hamiltonian matrix $H_{\rm hb}(\mathbf{k})$ at wavevector $\mathbf{k}$ is given as
\begin{eqnarray}
H_{\rm hb}(\mathbf{k})=
\begin{pmatrix}
0& \sum_{i=1}^{3}t_{i}\mathrm{e}^{-i\mathbf{k\cdot\tau_{i}}}\\
\sum_{i=1}^{3}t_{i}\mathrm{e}^{i\mathbf{k\cdot\tau_{i}}} &0
\end{pmatrix}.
\end{eqnarray}
where the subscript ``b'' means the bulk. 
$\boldsymbol{\tau}_{i=1,2,3}$ are expressed as $
\boldsymbol{\tau}_{1}=(0,1),
\boldsymbol{\tau}_{2}=(-\frac{\sqrt{3}}{2},-\frac{1}{2}),
\boldsymbol{\tau}_{3}=(\frac{\sqrt{3}}{2},-\frac{1}{2}),
$ and we put the length of the nearest-neighbor bonds as unity. For simplicity $t_i$ are assumed to be positive.
The primitive vectors $\mathbf{a}_{i=1,2}$ are
$
\mathbf{a}_1=(\frac{\sqrt{3}}{2},\frac{3}{2}),
\mathbf{a}_2=(-\frac{\sqrt{3}}{2},\frac{3}{2})
$, and the reciprocal primitive vectors are
$
\mathbf{G}_1=2\pi\frac{2}{3}(\frac{\sqrt{3}}{2},\frac{1}{2}),
\mathbf{G}_2=2\pi\frac{2}{3}(-\frac{\sqrt{3}}{2},\frac{1}{2}).
$

We first note that the bulk Hamiltonian $H_{\rm hb}$ has chiral symmetry: $\sigma_{z}H_{\rm hb}\sigma_{z}=-H_{\rm hb}$, where $\sigma_z$ is the Pauli matrix.
Therefore,
 if $|\psi\rangle$ is
an eigenstate with an eigenvalue $E$, $\sigma_{z}|\psi\rangle$ is an eigenstate with energy $-E$.
The eigenvalues are given by
\begin{eqnarray}
E_{\rm hb}(\mathbf{k})=\pm\left|t_i \sum_{i=1}^{3}\mathrm{e}^{-i\boldsymbol{\tau}_i\cdot\mathbf{k}}\right|.
\label{eq:egraphene}
\end{eqnarray}
Hereafter we put parameters as $t_2=t_3=1$, and $t_1 = t$, where $t$ is a real positive tunable parameter.
 The bulk dispersion is given as
\begin{eqnarray}
E_{\rm hb}^{2}
=\left(t+2\cos \frac{\sqrt{3}k_{x}}{2}\cos \frac{3k_y}{2}\right)^2  \nonumber\\  +4\cos^2 \frac{\sqrt{3}k_{x}}{2} \sin^2 \frac{3k_y}{2}.
\end{eqnarray}
Because of the chiral symmetry, the gap closes only at zero energy. 
The bulk gap-closing points $(k_x^{*},k_y^{*})$ are given by the equations: $\cos\frac{\sqrt{3}k_x^{*}}{2}=\pm \frac{1}{2} t$ and $\sin \frac{3}{2}k_y^{*}=0$.
The equations give two gap-closing points in the bulk BZ, and they exist for $t\leq 2$. 
The gap-closing points move with the change of the anisotropy $t$, as pointed out
in Ref.~\onlinecite{Dietl08}.
For $t=1$, i.e. the tight-binding model of graphene, the upper and lower bands touch at K $(\frac{2\pi\sqrt{3}}{9},\frac{2\pi}{3})$ and K' $(-\frac{2\pi\sqrt{3}}{9},\frac{2\pi}{3})$, and with the increase of $t$
the gap-closing points get closer along the line $k_{y}=\frac{2\pi}{3}$ (Fig.~\ref{fig:graphene-like}(b)). Around each of the two gap-closing points, the dispersion forms a Dirac cone, and Berry phase around each gap-closing point is $\pi$, which 
is protected by chiral symmetry. Because of this $\pi$ Berry phase, the gap-closing points do not disappear\cite{Okada10} as we change $t (<2)$.
The bulk gap-closing points move in the direction perpendicular to the bonds with anisotropic hopping integral $t$.
At $t=2$ the gap-closing points meet and they annihilate each other at $k_{x}=0$ (Fig.~\ref{fig:graphene-like}(b)) \cite{Dietl08}. This is possible because the sum of the Berry phase becomes zero, i.e. $\pi+\pi\equiv 0$ (mod $2\pi$). 
For $t > 2$, there are no bulk gap-closing points. 

The evolution of the edge states with the change of the anisotropy has been 
studied in several papers \cite{Kohmoto07,Dietl08,Delplace11}.
As we see in the following, for $t>2$ flat-band edge states on the zigzag or Klein edges completely cover the BZ, as has been studied in Ref.~\onlinecite{Delplace11}.
For the zigzag edges it occurs when the bond with hopping $t$ is perpendicular to the edge, 
and for the Klein edges it occurs when the bond with hopping $t$ is not perpendicular to the edge.
For these cases with zigzag and Klein edges, 
we calculate dispersions in Fig.~\ref{fig:graphene-like} (c) for $t_1=1$ and $t_1=2.2$, at $t_2=t_3=1$ in both cases. 
The flat-band edge states are separated completely from the bulk for $t_1=2.2$ (Fig.~\ref{fig:graphene-like}(c)).

To explain this behavior, we solve
the Schr$\ddot{\mathrm{o}}$dinger equation 
in the semi-infinite geometry with a zigzag edge. The zigzag edge 
is assumed to be perpendicular to the bonds with hopping integral $t_1$.
We express the wavefunction $|\Psi^{ }(k)\rangle
$ as
\begin{eqnarray}
|\Psi^{ }(k)\rangle =\sum_{i=1} (a^{ }_{i}(k)|A^{ }_{i}(k)\rangle+ b^{ }_{i}(k)|B^{ }_{i}(k)\rangle),
\end{eqnarray}
 where $i$ denotes an index for unit cells containing two sublattice sites, A and B, counted from the edge $(i=1)$,
 $k$ is the wavenumber along the edge, and $a^{ }_{i}(k)$ ($b^{ }_{i}(k)$)
 denotes the coefficient for the wavefunctions at A(B) sublattice, $|A^{ }_{i}(k)\rangle$
($|B^{ }_{i}(k)\rangle$).
Acting $H_{\rm h}$ onto $|A^{ }_{i}(k)\rangle$ and $|B^{ }_{i}(k)\rangle$, we have
\begin{eqnarray}
&&\langle B^{ }_{i}(k) |H_{\rm h}|A^{ }_{i}(k)\rangle =t_1,\\
&&\langle B^{ }_{i-1}(k) |H_{\rm h}|A^{ }_{i}(k)\rangle =t_2+t_3\mathrm{e}^{-ik}.
\end{eqnarray}
From Fig.~\ref{fig:graphene-like}(c), the surface states are expected to 
be at the zero energy, and as we see later it is the case indeed. When we set 
the eigenvalue to be zero, $H_{\rm h}|\Psi(k)\rangle=0$, 
we obtain
\begin{eqnarray}
a^{ }_{i}(t_2+t_3\mathrm{e}^{-ik})+a^{ }_{i+1}t_1=0,&
b^{ }_{i}=0.
\end{eqnarray}
Thus the amplitude of the flat-band states is given by \begin{eqnarray}
a^{ }_{n}(k)=a^{ }_{1}\left[-\frac{t_2+t_3\mathrm{e}^{-ik}}{t_1}\right]^{n-1}
,&
b^{ }_{i}=0
\label{eq:AB}
\end{eqnarray} where $a^{ }_{1}=\left[1-\frac{|t_2+t_3\mathrm{e}^{-ik}|^2}{t^2_1}\right]^{-1/2}$ from normalization, 
and
the condition for existence of the edge states, i.e. normalizability of the wavefunction, is given as
\begin{eqnarray}
\left|\frac{t_2+t_3\mathrm{e}^{-ik}}{t_1}\right|< 1.
\label{eq:grapheneedgecond}
\end{eqnarray}
For example, for $t_1=t_2=t_3$ (graphene model), the wavenumber that satisfies the
condition (Eq.~(\ref{eq:grapheneedgecond})) for existence of the edge state is given as $\frac{2\pi}{3}<k<\frac{4\pi}{3}$,
which agrees with the well-known flat band in graphene ribbon with a zigzag edge \cite{Fujita96}.
In addition, by the relation
$
\left|\frac{t_2+t_3\mathrm{e}^{-ik}}{t_1}\right| <\frac{t_2+t_3}{t_1},
$
when $t_1>t_2+t_3$ is satisfied, the wavefunction defined by 
(\ref{eq:AB}) is normalizable for every $\mathbf{k}$ and
the flat bands cover the whole 1D BZ . This condition $t_1>t_2+t_3$ means that
the anisotropy is sufficiently large.
These results agree with numerical calculations in Fig.~\ref{fig:graphene-like} (c).

In Fig.~\ref{fig:graphene-like} (c) we also show results for armchair edges and for 
Klein (bearded) edges. For armchair edges there are no flat-band edge states.
For Klein edges where ${\tau}_1$-bonds (hopping $t_1$) are not perpendicular to the edge, there are flat-band edge states; if $t_1>2$ the flat-band edge states cover 
the entire BZ. 
These results agree with the results in Ref.~\onlinecite{Delplace11}.

\section{Diamond lattice}\label{Sec:Diamond}
\begin{figure}[htbp]
 \begin{center}
 \includegraphics[width=80mm]{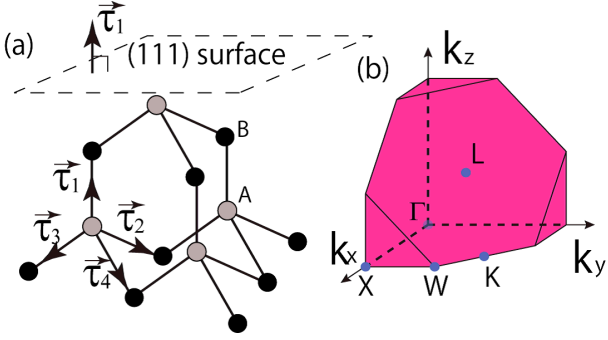}
 \caption{(a) Schematic of a diamond lattice. A (gray) and B (black) atoms denote the two sublattices. $\boldsymbol{\tau}$s 
represent vectors connecting nearest-neighbor atoms.
(b) The BZ of the diamond lattice structure within $k_{x,y,z} >0$. }
\label{fig:DLattice}
\end{center}\end{figure}

In the previous section, we have seen that the flat-band edge states of the honeycomb-lattice model with the large anisotropy cover the whole BZ.
In the similar way as in the honeycomb-lattice model, 
in this section we show surface flat bands
in the tight-binding model on the
 diamond lattice.
 The existence of such a flat-band surface states has been 
proposed in Ref.~\onlinecite{Takagi08} for an isotropic case. 
In the following we extend this concept to anisotropic cases
and study properties of the wavefunctions. 
 In the honeycomb lattice in the previous section, 
the flat-band edge states exist between certain wavenumbers, which 
are identified with the bulk gap-closing points.
 We will show similar phenomena for the diamond lattice.
The Hamiltonian $H_{\rm d}$ is 
\begin{eqnarray}
H_{\rm d}=\sum_{\langle ij \rangle}c^{\dagger}_{i}t_{ij}c_{j},
\label{eq:TB}
\end{eqnarray}
where $t_{ij}=t_{ji}$ is the hopping integral from site $i$ to $j$, and the suffix d means the diamond lattice.
We assume that the hopping integrals $t_{ij}$ are real positive parameters. The nearest-neightbor bond vectors, $\boldsymbol\tau$s, are as follows: $\boldsymbol\tau_1=\frac{1}{4}(1,1,1)$, $\boldsymbol\tau_2=\frac{1}{4}(-1,1,-1)$, $\boldsymbol\tau_3=\frac{1}{4}(-1,-1,1)$, and $\boldsymbol\tau_4=\frac{1}{4}(1,-1,-1)$ (Fig.~\ref{fig:DLattice}).
The four hopping integrals are labeled with $\boldsymbol{\tau}_{a}$ as $t_{\boldsymbol{\tau}_{a}}=t_{a}$.
The bulk Hamiltonian matrix $H_{\rm db}(\mathbf{k})$ is represented as
\begin{eqnarray}
H_{\rm db}(\mathbf{k})=
\begin{pmatrix}
0& \sum_{i=1}^{4}t_{i}\mathrm{e}^{-i\mathbf{k\cdot\boldsymbol\tau_{i}}}\\
\sum_{i=1}^{4}t_{i}\mathrm{e}^{i\mathbf{k\cdot\boldsymbol\tau_{i}}} &0
\end{pmatrix}.
\end{eqnarray}
This also preserves the chiral symmetry, $\sigma_{z}H_{\rm db}\sigma_{z}=-H_{\rm db}$, and therefore the eigenenergy 
is symmetric with respect to $E\leftrightarrow -E$.
As is proposed in Ref.~\onlinecite{Takagi08}, when $t_{1-4}$ are identical, the surface states on the (111) surface 
for this Hamiltonian forms a flat band, which partially cover the surface BZ. 

\begin{figure}[htbp]
 \begin{center}
 \includegraphics[width=80mm]{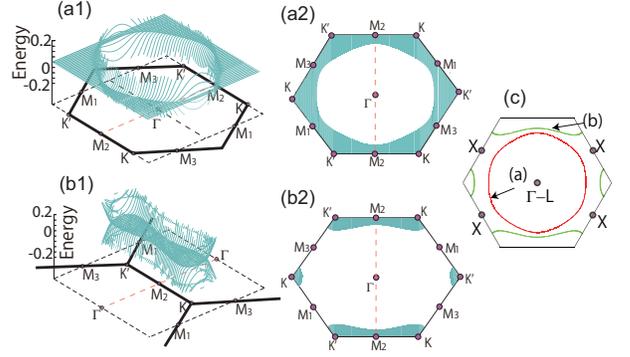}
 \caption{
 Band structure for the diamond lattice model with the $(111)$ surface, for the two cases, 
(a) $t_1=t$, $t_2=t_3=t_4=1$ and (b) $t_3=t$, $t_1=t_2=t_4=1$.
(a1)(b1) show the respective band structures close to zero energy for (a)(b). The thick lines represent the 
first BZ. (a2)(b2) show the region with the surface flat bands at zero energy. In (c), the FS at zero energy in the 3D bulk BZ projected on the $(111)$ plane for the cases (a) and (b). }
\label{fig:FSB}
\end{center}\end{figure}

\begin{figure}[htbp]
 \begin{center}
 \includegraphics[width=80mm]{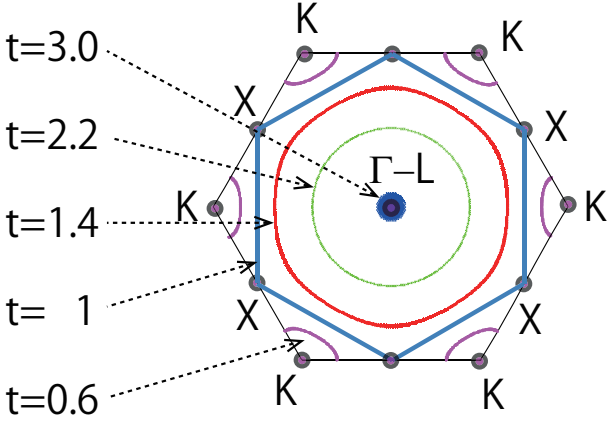}
 \caption{The FSs at zero energy in the 3D bulk BZ projected on the $(111)$ plane for $t_2=t_3=t_4=1$ and $t_1 =t=0.6,1, 1.4, 2.2, 3.0$ are shown.
For $t>1$, the loop shrinks with increase in $t$, and at $t=3.0$ the loop becomes a dot.}
\label{fig:sunflower}
\end{center}\end{figure}

First, we set one of the four hopping integrals as a tunable positive parameter $t$ and the other hopping integrals as unity.
For $t=1.4$, the dispersions for slab geometry with $(111)$ surface are shown in Fig.~\ref{fig:FSB}.
We show the results for cases, (a) $t_1=t$, $t_2=t_3=t_4=1$ and (b) $t_3=t$, $t_1=t_2=t_4=1$, whose differences
lie in their surface orientation relative to the anisotropy; namely, the stronger bond is 
perpendicular to the surface for (a) and not perpendicular for (b). 
Figures~\ref{fig:FSB} (a1) and (b1) show the dispersions for the slab geometry near zero energy for (a) and (b), respectively. As one can see, there is a surface flat band at zero 
energy, which partially covers the BZ. 
The distributions of zero-energy flat bands in the surface BZ are shown in (a2)(b2).
On the other hand, 
Fig.~\ref{fig:FSB}(c) shows the projections of the bulk 
gap-closing points at $E=0$ onto the $(111)$ surface for (a)(b).
Notably, it is a novel property of the model that the gap closes along a loop
in $\mathbf{k}$ space.
This comes because $H_{\rm db}$ consists only of $\sigma_x$ and $\sigma_y$, 
and one should tune only two parameters (the coefficients of $\sigma_x$ and
$\sigma_y$) to be zero to close the gap. Such gap-closing points form a loop in the bulk BZ.
In particular, if one sets the Fermi energy to be zero, the gap-closing loop 
becomes the Fermi surface (Fermi loop). 
By comparing the bulk FSs in Fig.~\ref{fig:FSB}(c) with the flat-band surface 
states (a)(b), we see the correspondence between the bulk FS and the distribution of the flat surface bands, as is discussed in the previous case of the honeycomb lattice; the verge of the flat surface bands corresponds to the bulk FS projected onto the surface direction. 
 In this case, the surface dispersion is determined by the relative orientation between the nearest-neighbor 
hopping vector with $t$ and the surface.
 Therefore by rotating the surface orientation in (b) from $(111)$ to $(\bar1\bar11)$, we have the same flat-band states as shown in (a). 
From these observations, one can see how an anisotropy affects the surface flat band. In particular, if $t_1$ is increased, the surface flat band gradually grows within the surface BZ and eventually cover the 
whole BZ, as we see in the following. 
The bulk dispersion for $t_{1}=t$, $t_2=t_3=t_4=1$ is given by
\begin{eqnarray}
E_{\rm db}^{2}
&=&\left|\sum_{i=1}^{4}t_{i}\mathrm{e}^{-i\mathbf{k}\cdot\boldsymbol{\tau}_{i}}\right|^2\nonumber\\
&=&\left(t+\cos\frac{k_x+k_y}{2}+\cos\frac{k_y+k_z}{2}+\cos\frac{k_z+k_x}{2}\right)^2\nonumber\\ 
&\ &+\left(\sin\frac{k_x+k_y}{2}+\sin\frac{k_y+k_z}{2}+\sin\frac{k_z+k_x}{2}\right)^2.\nonumber\\ \label{eq:bFS}
\end{eqnarray}
Therefore, the gap-closing points $E_{\rm bd}=0$ are given by two equations,
$t+\cos\frac{k_x+k_y}{2}+\cos\frac{k_y+k_z}{2}+\cos\frac{k_z+k_x}{2}=0$ and
$\sin\frac{k_x+k_y}{2}+\sin\frac{k_y+k_z}{2}+\sin\frac{k_z+k_x}{2}=0$ (Fig.~\ref{fig:sunflower}). The FS encircles the $\Gamma$-$L$ line in the bulk BZ for $t\geq 1$.
This can be explicitly seen when the wavevector is close to the $L$ point, $\mathbf{k}\sim (\pi,\pi,\pi)$, which is true when $t$ is close to $3$.
Around the $L$ point, by putting $\mathbf{k}=(\pi,\pi,\pi)+(\delta k_x,\delta k_y,\delta k_z)$, the FS loop is expressed as
\begin{eqnarray}
(\delta k_x)^2+(\delta k_y)^2+(\delta k_z)^2=8(3-t),&
\delta k_x+ \delta k_y+ \delta k_z=0 \nonumber\\
\end{eqnarray}
 from Eq.~(\ref{eq:bFS}). Thus, the FS is a circle of radius $\sqrt{8(3-t)}$, surrounding the $L$ points. 
 We see that the FS is getting smaller with the increase of $t$ and shrinks to the $\Gamma$ point at $t= 3$, while the flat-band surface states expand with $t$.
For $t > 3$ the loop vanishes, and the surface flat band covers the whole surface BZ.

\subsection{Flat-band states localized at the surface}
Here we calculate the Bloch wavefunction of the flat-band surface states for the tight-binding model (Eq.~(\ref{eq:TB})) with the (111) surface. To this end, we consider a semi-infinite geometry with the $(111)$ surface.
We derive a wavefunction with zero eigenvalue as
$|\phi\rangle=\sum_{i=1}(a_{i}(\mathbf{k})|A_{i}(\mathbf{k})\rangle+b_{i}(\mathbf{k})|B_{i}(\mathbf{k})\rangle)$,
 where $i$ denotes indices of the unit cell counted
 from the surface $(i=1)$, $|A_i(\mathbf{k})\rangle$ ($|B_i(\mathbf{k})\rangle$)
denotes the wavefunction at the $i$th layer with the wavevector $\mathbf{k}$, and $a_{i}(\mathbf{k})(b_{i}(\mathbf{k}))$
 denotes the coefficient of the wavefunction for each sublattice.
The matrix element of the Hamiltonian $H_{\rm d}$ is given as
\begin{eqnarray}
&&\langle B_{j-1}(\mathbf{k})| H_{\rm d}|A^{}_{j}(\mathbf{k})\rangle=t_1, \\
&&\langle B_{j}(\mathbf{k})| H_{\rm d}|A_{j}(\mathbf{k})\rangle=t_2 \mathrm{e}^{-ik_1}+t_3+t_4\mathrm{e}^{ik_2},
\end{eqnarray}
where $k_{1}=\mathbf{k}\cdot(\boldsymbol{\tau}_{2}-\boldsymbol{\tau}_{3})$ and $k_{2}=\mathbf{k}\cdot(\boldsymbol{\tau}_{3}-\boldsymbol{\tau}_{4})$.

In order to calculate the flat-band surface states, 
we impose the state $|\phi\rangle$ 
to have zero energy $H_{\rm d}|\phi \rangle=0$.
In the Schr$\ddot{\mathrm{o}}$dinger equation, $H_{\rm d}|\phi \rangle$ can be calculated as
\begin{eqnarray}
H_{\rm d}|\phi(\mathbf{k})\rangle&=&
\sum_{j=2}^{\infty}[a_{j}(t_2 \mathrm{e}^{-ik_1}+t_3+t_4\mathrm{e}^{ik_2} )+a_{j+1}t_{1}]|B_{j}(\mathbf{k})\rangle
\nonumber\\
&+&
[b_{1}(t_2 \mathrm{e}^{-ik_1}+t_3+t_4\mathrm{e}^{ik_2} )]|A_{1}(\mathbf{k})\rangle\nonumber\\
&+&\sum_{j=2}^{\infty}
[b_{j-1} t_{1}+b_{j}(t_2 \mathrm{e}^{ik_1}+t_3+t_4\mathrm{e}^{-ik_2} )]|A_{j}(\mathbf{k})\rangle,\nonumber\\
\end{eqnarray}
From $H_{\rm d}|\phi\rangle=0$, 
the amplitudes are derived as $b_i(k)=0$, and
\begin{eqnarray}
a_{j+1}(\mathbf{k})=-\frac{t_2\mathrm{e}^{-ik_1}+t_3+t_4\mathrm{e}^{ik_2}}{t_1}a_{j}(\mathbf{k}).
\label{eq:samplitude}
\end{eqnarray}
Namely, the zero-energy states localize near the surface for every wavevector if $t_1>t_2+t_3+t_4$, 
because from Eq.~(\ref{eq:samplitude}) we have
\begin{eqnarray}
\left|\frac{a_{j+1}(\mathbf{k})}{a_{j}(\mathbf{k})}\right|
=\left|\frac{t_2\mathrm{e}^{-ik_1}+t_3+t_4\mathrm{e}^{ik_2}}{t_1}\right|
\nonumber\\
\leq \frac{t_2+t_3+t_4}{t_1}<1.
\label{eq:seqamplitude}
\end{eqnarray}
The second equality in Eq.~(\ref{eq:seqamplitude}) holds only at the $\Gamma$ point in the 2D surface BZ.
In addition, from Eq.~(\ref{eq:seqamplitude}) a penetration depth $\lambda$, defined as $|a_j|\sim e^{-j/\lambda}$, is given by $\mathrm{e}^{-\frac{1}{\lambda}}=\left|\frac{t_2\mathrm{e}^{-ik_1}+t_3+t_4\mathrm{e}^{ik_2}}{t_1}\right|$. Hence the penetration depth of the surface states is maximum at $\mathbf{k}=0$. As we see in the following, 
the longest penetration depth at $\mathbf{k}=0$ means that 
the finite-size effect is largest there.

\begin{figure}[htbp]
 \begin{center}
 \includegraphics[width=90mm]{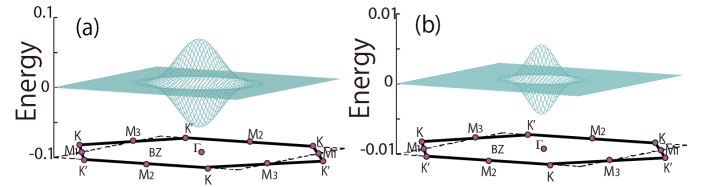}
 \caption{Dispersion near zero energy at $t=3.4$ for the slab geometry. 
The number $N$ of unit cells to the direction normal to the surface is
$N=20$ in (a) and $N=40$ in (b).}
\label{fig:SCFB}
\end{center}\end{figure}

From the above discussion, the model has the completely flat band at zero energy when the system is semi-infinite.
However, if the thickness of the slab is finite, 
there will be a small splitting of energy to $\pm\Lambda$ due to hybridization of the surface states at the opposite
sides of the slab. 
This gap is larger when the penetration depth is longer. Therefore, 
the gap is expected to be largest at the $\Gamma$ point. 
We calculate the band structure for 
a slab geometry
in Fig.~\ref{fig:SCFB}(a)(b), and see that it is indeed the case.
The maximum value of the finite-size gap $\Lambda$ is estimated as the following.
We focus on the $\Gamma$ point, and assume $(t_2+t_3+t_4)/t_1=\alpha<1$.
When the thickness of the slab $N$ is large, the surface wavefunction for the top surface $|\phi^{t}\rangle$
 and that for the bottom surface $|\phi^{b}\rangle$
can be treated separately.
 The wavefunctions are approximately given as
\begin{eqnarray}
|\phi^{t}(\mathbf{k})\rangle\sim \sum_{n=1}^{N}ac^{n-1}|A_{i}(\mathbf{k})\rangle,\ 
|\phi^{b}(\mathbf{k})\rangle\sim \sum_{n=1}^{N}ac^{N-n}|B_{i}(\mathbf{k})\rangle, \nonumber\\
\end{eqnarray}
where $c=-(t_2\mathrm{e}^{-ik_1}+t_3+t_4\mathrm{e}^{ik_2})/t_1$, and $a^2=(1-|c|^2)/(1-|c|^{2N})$.
Then we obtain the hybridization as
\begin{eqnarray}
\langle \phi^{b}(\mathbf{k})|H|\phi^{t}(\mathbf{k})\rangle\sim a^2c^{N}t_1, 
\end{eqnarray}
which is expected to give the size of the gap due to the finite-size effect.
At the $\Gamma$ point, $|c|$ becomes maximum and therefore the finite-size effect of the energy is largest at $\Gamma$, 
taking its maximum value 
\begin{equation}
\Lambda \sim t_1(1-\alpha^{2})\alpha^{N}.
\label{eq:Lambda}
\end{equation}
For $t_1=3.4$, $t_2=t_3=t_4=1.0$, Eq.~(\ref{eq:Lambda}) gives $\Lambda \sim 0.0616$ for $N=20$, and $\Lambda \sim 0.00504$ for $N=40$. 
On the other hand, our band-structure calculation (Fig.~\ref{fig:SCFB})
gives $\Lambda$ to be 
$\Lambda \sim 0.0632$ for $N=20$, and $\Lambda=0.005042$ for $N=40$. Thus our estimate for $\lambda$ in Eq.~(\ref{eq:Lambda}) well agrees with the numerical calculation, showing that the gap around $\mathbf{k}=0$ is governed by the
penetration depth of the surface states into the bulk. 

\subsection{Topological transition of the bulk FS}
\begin{figure}[htbp]
 \begin{center}
 \includegraphics[width=80mm]{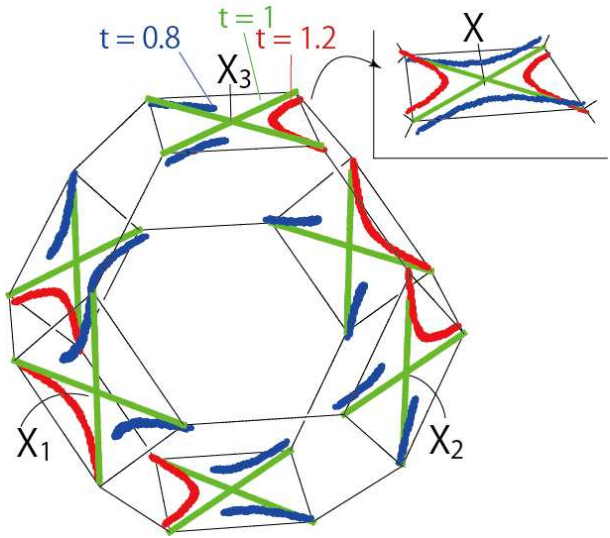}
 \caption{The FSs in the 3D bulk BZ for $t_2=t_3=t_4=1$ and $t_1 =t=$0.8, 1, 1.2 are shown. 
At $t =$0.8 (blue) the FSs consist of two open orbits. On the other hand at $t =$ 1.2 (red), the FS becomes one closed orbit in the BZ.
To illustrate the topological transition of the FS at $t = 1$ (green), 
the FSs around one of the X points are magnified in the inset in the 
extended zone scheme.}
\label{fig:t1_08_10_12}
\end{center}\end{figure}

As can be seen in Fig.~\ref{fig:sunflower}, the topology of the FS loop changes at $t=1$. In this section, we study this topological transition of the FS in the diamond-lattice model.
We assume that $t_{2,3,4}$ are fixed to be unity, whereas $t_1=t$ is varied. 
The FSs for $t=1.2,1,0.8$ are shown in Fig.~\ref{fig:t1_08_10_12}.
When $t<1$ there are two FS loops forming open orbits, though it is not
immediately seen in Fig.~\ref{fig:t1_08_10_12}. On the other hand, 
when $t=1.2$ there is a single FS loop forming a closed orbit. 
At $t=1$ the topology of the
FS changes at the three X points. 
This change at the X points is not clearly seen, 
because the X points are on the BZ boundary.
To clarify the topology change, in the inset of Fig.~(\ref{fig:t1_08_10_12}) 
we show the FSs close to the X points in the
extended zone scheme.
As $t$ is increased further, 
at $t=3$ the FS shrinks to a point and simultaneously the surface flat band extends cover the whole BZ.

\section{Topological explanation for existence of the flat-band states}
In Ref.~\onlinecite{Ryu02}, a topological interpretation of the 
existence of edge states at zero energy in two-dimensional models with 
chiral symmetry is given. In this section we apply this theory to the 
present models on the honeycomb lattice and the diamond lattice, and 
show that the edge/surface flat band states are explained within this 
theory \cite{Ryu02}.

To apply the topological argument in Ref.~\onlinecite{Ryu02}, 
the crystal termination is crucial. 
The way how the edges are oriented and how the crystal is terminated is
incorporated into the formalism in the following way. 
For two-dimensional models with chiral symmetry, for example, we begin with a bulk system, 
and we cut the system along one direction by cutting the nearest-neighbor 
bonds, in order to discuss edge states. Let $y$ denote the coordinate along which the system will be cut.
Then, 
following Ref.~\onlinecite{Ryu02}, we expand the bulk Hamiltonian by the Pauli matrices $\sigma_{x},\sigma_{y}$ as $H=h_{x}(k'_x,k_y)\sigma_x+h_{y}(k'_x,k_y)\sigma_y$. Here 
$k_y$ denotes the component of the wavevector 
along the $y$-direction (along the edge), 
and $k'_x$ is the other component of the wavevector. We note that because we
assume chiral symmetry $\sigma_z H\sigma_z=-H$, 
the 2$\times$2 Hamiltonian $H$ has no $\sigma_z$ term.
Because the bulk system is cut along the $y$-axis, $k'_x$ will no longer be a good
quantum number. Then the criterion in Ref.~\onlinecite{Ryu02} says 
that if the trajectory of $(h_x,h_y)$ for the change of $k'_x$ with fixed $k_y$ 
encircles the origin, zero-energy edge states exist for the given $k_y$. If not, 
zero-energy edge states will not exist \cite{Ryu02}. 
An intuitive picture of this argument is the following. The origin $(h_x,h_y)=(0,0)$ is a singular point because the bulk 
Hamiltonian has degenerate eigenvalues at zero energy. Whether the trajectory
encircles this singularity or not determines a classification of the Hamiltonian 
either into a class with no edge state or a class with flat-band edge states. Namely, if the trajectory does not encircle the origin, it can be continuously 
deformed into a point without encountering the singular point, which leads to an
absence of zero-energy boundary states.
\begin{figure}[htbp]
 \begin{center}
 \includegraphics[width=80mm]{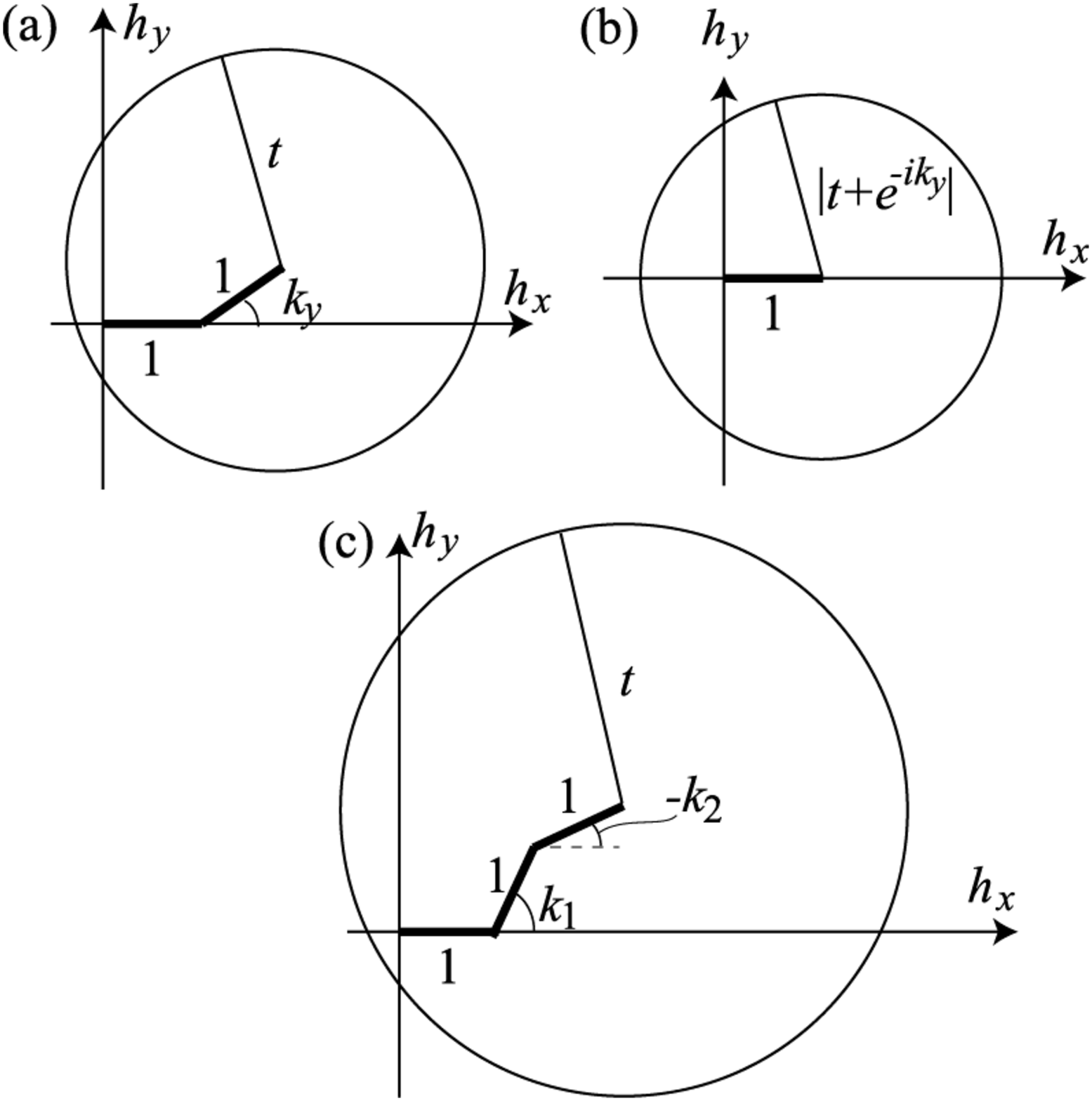}
 \caption{Trajectories of $(h_{x},h_{y})$ by varying $k'_x$ in (a) (b) or $k_3$ in (c) for 
the honeycomb-lattice model with zigzag (a), with Klein (b) edges,
 and in the diamond-lattice model with the $(111)$ surface (c). In these trajectories
 $k_y$, $k_1$ and $k_2$ are fixed.
}
\label{fig:topo-edge}
\end{center}\end{figure}

We apply this criterion to the present models to show that 
the flat-band boundary states discussed so far are fully explained by this theory.
For the anisotropic honeycomb-lattice models, explanations
are given in Ref.~\onlinecite{Delplace11}, and we reproduce it here for illustration. For zigzag edges we have
\begin{eqnarray}
&&h_x=t\cos(k_y-k'_x)+1+\cos k_y,\\
&&h_y=-t\sin(k_y-k'_x)+\sin k_y.
\end{eqnarray}
Hence the trajectory is a circle with a radius $|t|$ centered at $(1+\cos k_y,\sin k_y)$ (Fig.~\ref{fig:topo-edge}(a)). The condition that it encircles the origin reproduces the range of the wavevector of 
the flat-band edge states, obtained in the previous section. In particular, for $t>2$ the trajectory encompasses the
origin irrespective of the value of $k_y$, and existence of 
the perfectly flat edge band over the whole BZ results, as we discussed previously.
The case for the Klein edge is explained similarly, where we have
\begin{eqnarray}
&&h_x=1+\cos(k_y-k'_x)+t\cos k'_x,\\
&&h_y=-\sin(k_y-k'_x)+t\sin k'_x.
\end{eqnarray}
with the trajectory shown in Fig.~\ref{fig:topo-edge}(b). Then it is easily seen that the flat-band 
edge states extend over the whole BZ when $t>2$. 

So far the topological characterization of flat-band edge
states is only for edge states in two-dimensional systems 
\cite{Ryu02}. We can extend this discussion to three-dimensional models such 
as our diamond-lattice model. For this model we obtain
\begin{eqnarray}
&&h_x=1+\cos k_1+\cos k_2+t\cos k_3,\\
&&h_y=\sin k_1-\sin k_2-t\sin k_3,
\end{eqnarray}
where $k_1={\bf k}\cdot(\boldsymbol{\tau}_2-\boldsymbol{\tau}_3)$,
$k_2={\bf k}\cdot(\boldsymbol{\tau}_3-\boldsymbol{\tau}_4)$, and
$k_3={\bf k}\cdot(\boldsymbol{\tau}_3-\boldsymbol{\tau}_1)$. 
When we cut the bonds parallel to $\boldsymbol{\tau}_1$, $k_3$ will no 
longer be a good quantum number. Then, by extending the argument in 
Ref.~\onlinecite{Ryu02} we conclude the following. If the trajectory of $(h_x,h_y)$ by 
the change of $k_3$ encircles the origin, there should be zero-energy edge state.
For $t>3$, it holds true irrespective of the values of $k_1$ and $k_2$, and therefore for $t>3$
the surface flat band covers the surface BZ (Fig.~\ref{fig:topo-edge}(c)).

This theory in Ref.~\onlinecite{Ryu02} also explains the reason why in these models 
the verge of the edge/surface states should be identical with the projection of 
the bulk gap-closing points/curves. 
Let $\mathbf{k}^{*}$ denote a wavenumber where 
the bulk eigenenergy is zero, i.e. $h_{x}=0$ and $h_{y}=0$. Because of the 
chiral symmetry, at such point $\mathbf{k}^{*}$ the bulk band gap is closed. 
In our honeycomb-lattice model there are two $\mathbf{k}^{*}$ points (Fig.~\ref{fig:graphene-like}). The set of $\mathbf{k}^{*}$ points in the diamond lattice model forms a loop (Figs.~\ref{fig:FSB} and \ref{fig:sunflower}). At such $\mathbf{k}^{*}$ points 
the trajectory of $(h_x,h_y)$ goes across the origin, and therefore 
it should be on the boundary between the regions where the edge/surface flat band 
exist or not. Therefore, the projections of the bulk zero-energy points onto the 
edge/surface are identical with the verge of the flat-band edge/surface states, 
which is one of the consequences in Ref.~\onlinecite{Ryu02}.
To summarize, this topological notion enables us to show existence or absence 
of flat-band
boundary states for various models with various boundary conditions.
The flat-band edge/surface states vary by changing anisotropy of the hopping integral.
We emphasize here that the existence of flat-band edge/surface states here does not result from 
neither interaction nor isolated atomic orbitals, but from topological structure in ${\bf k}$ space.

\section{Completely localized edge/surface states}
\label{sec:spatiallylocalizedstates}
We have found that nearest-neighbor tight-binding models on the bipartite lattices, such as the honeycomb and the diamond lattices,
 have flat-band boundary states covering the whole BZ, when the anisotropy of their hopping integrals is sufficiently large.
In general, when systems have completely flat bands over the entire BZ, one can construct a wavefunction which is spatially localized on a finite number of sites. 
Namely, because of the flatness of the band, any linear combination of the
eigenstates within this flat band is also an eigenstate; therefore by taking
an appropriate linear combination, one can construct a fully localized state.
This is analogous to constructing a spatially localized state as a linear combination of plane waves. 

\begin{figure}[htbp]
 \begin{center}
 \includegraphics[width=80mm]{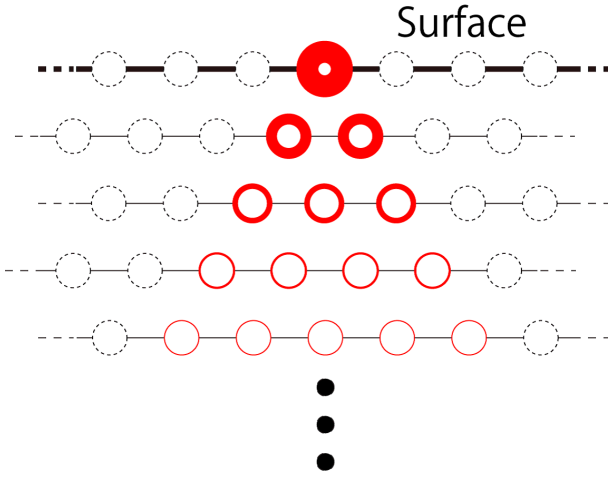}
 \caption{
Schematic of spatial distribution of the flat-band surface states.
The circles are the A sublattice sites, while the amplitudes on B sublattices are zero and are omitted.
The dotted circles show that their amplitudes of the wavefunction are zero.
The line thickness of the circle shows the magnitude of the amplitude.
The top atom has the largest amplitude, and the distribution of the wavefunction spatially spreads toward the interior with exponential decay.
This picture of the fully localized states applies both to the honeycomb lattice with the zigzag edge and to the diamond lattice with the (111) surface.}
\label{fig:SurfaceLocalizedtates}
\end{center}\end{figure}

The construction of fully localized state 
is possible only when the flat band covers the whole BZ. 
In this section, we calculate the fully localized wavefunction in the present models.
This wavefunction is exponentially decaying in the direction normal to the boundary, while on the outermost atomic 
layer, the wavefunction is nonzero only on a single site, as schematically shown in Fig.~\ref{fig:SurfaceLocalizedtates}. We consider semi-infinite systems for the honeycomb lattice with the zigzag edge and those for the diamond lattice with the $(111)$ surface. 
Similarly to the previous sections, the outermost atomic layer are assumed to belong
to the A sublattice.

\begin{figure}[htbp]
 \begin{center}
 \includegraphics[width=70mm]{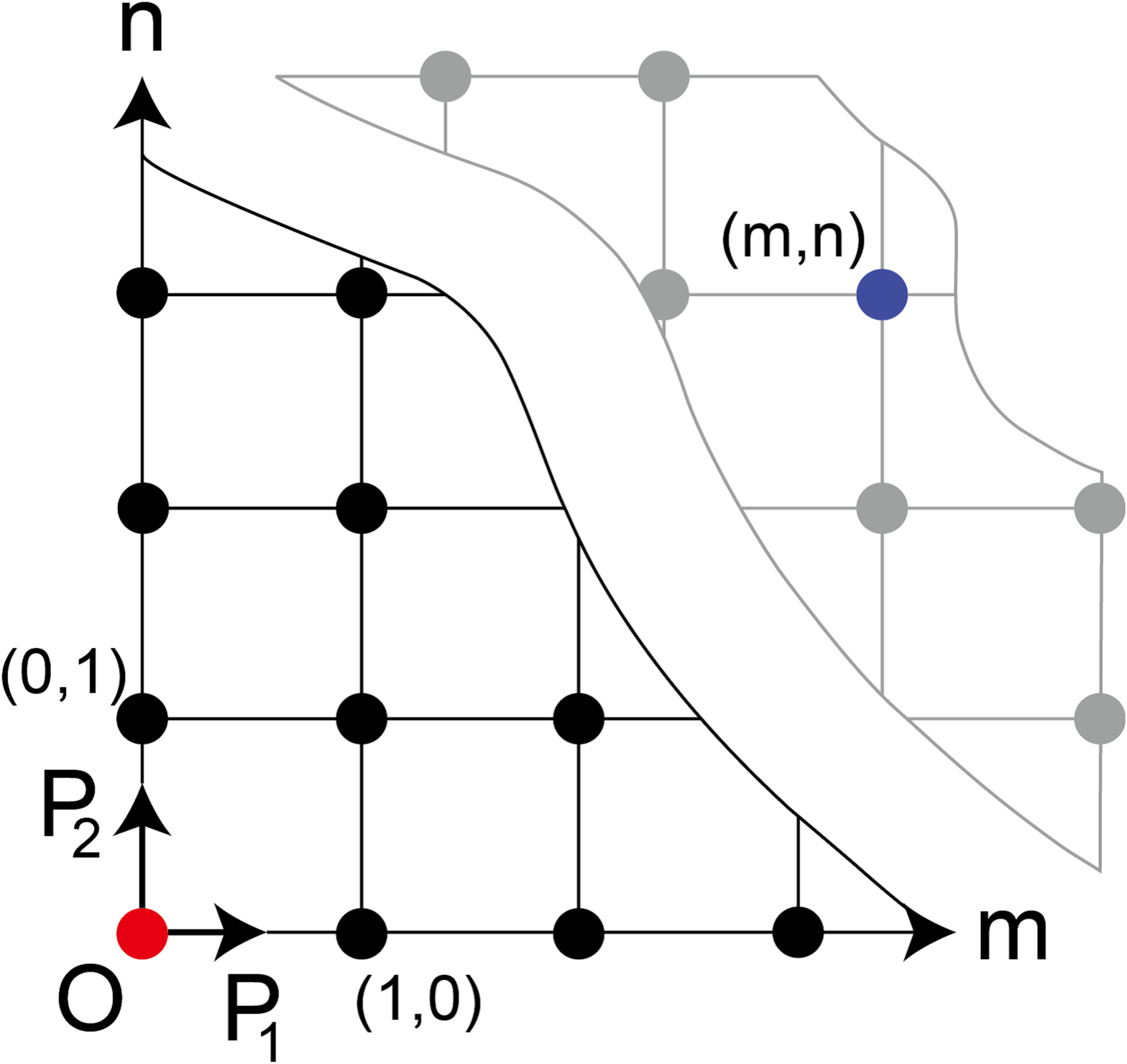}
 \caption{
Schematic of the grids for the shortest path problem.
The mover is at first at the origin $\mathcal{O}$, and it moves along $\mathbf{e}_{i}$ with the probability $P_{i}$, where $(\mathbf{e}_{i})_{j}=\delta_{ij}$.
}
\label{fig:PProblem}
\end{center}\end{figure}

We first consider the honeycomb lattice in the half plane $y\leq 0$. The zigzag edge is along the $x$ axis, and the origin $\mathcal{O}$ is set to be one site on the zigzag edge. 
The state at the site $-m \mathbf{a}_{1}-n\mathbf{a}_{2}$ is denoted as $|A_{mn}\rangle$ where $\mathbf{m}=(m, n)$
are nonnegative integers, and $\mathcal{O}=(0, 0)$.
We assume that the amplitude of the localized wavefunction on the outermost atomic layer is nonzero only at $\mathcal{O}$. 
We express the wavefunction of the localized states as
\begin{eqnarray}
|\Psi^{ }\rangle=\sum_{m,n}a_{mn} |A_{mn}\rangle,
\end{eqnarray}
 where $a_{mn}$ is the amplitude for $|A_{mn}\rangle$ at A sublattice.
 The amplitude at B sublattice is identically zero.
From the Schr$\ddot{\mathrm{o}}$dinger equation, we have a relation between amplitudes as
\begin{eqnarray}
a_{mn} =-\frac{t_{2}}{t_1}a_{m-1n}-\frac{t_{3}}{t_1}a_{mn-1}.
\label{eq:t1a}
\end{eqnarray}
Intriguingly, 
the solution for the above sequence determined by (\ref{eq:t1a}) 
is the same as the following problem.
Consider a mover in the $xy$ plane on the grid shown in Fig.~\ref{fig:PProblem}.
The mover is first on the $\mathcal{O} = (0,0)$ site. At each step it moves by $(1,0)$ with a probability $P_{1}$, by $(0,1)$ with a probability $P_{2}$ (Fig.~\ref{fig:PProblem}), and the movement is finished otherwise.
Finally after $m+n$ steps the probability $P_{mn}$ that the mover is at $(m,n)$ ($m,n \geq 0$) along the shortest paths is given as
\begin{eqnarray}
P_{mn}=\frac{(m+n)!}{m!n!}P_{1}^{m}P_{2}^{n}.
\end{eqnarray}
By replacing the probabilities of the movement with the ratio of the hopping integrals, 
\begin{eqnarray}
P_{i}\to -\frac{t_{i+1}}{t_1},
\end{eqnarray}
where $i=1,2$, we have the amplitude of the wavefunction as
\begin{eqnarray}
a_{mn}=\frac{1}{Z_{\rm h}} \frac{(m+n)!}{m!n!}\left(-\frac{t_{2}}{t_1}\right)^{m} \left(-\frac{t_{3}}{t_1}\right)^{n}\ (m,n\geq 0),
\end{eqnarray}
where $Z_{\rm h}$ is the normalization constant. For $m<0$ or $n<0$, $a_{mn}$ vanishes.
In this spatial representation of the wavefunction, the condition for existence of the fully localized states on the boundary is that the wavefunction is normalizable.
Generally, because $t_{1,2,3}$ are positive, $Z_{\rm h}$ satisfies the following relation:
\begin{eqnarray}
Z_{\rm h}^2&=&\sum_{N=0}^{\infty}\sum_{n,m=0}^{N}\delta_{n+m,N}P^{2}_{mn}\nonumber\\
& \leq &\left(\sum_{N=0}^{\infty} \sum_{n,m=0}^{N=n+m}|P_{mn}|\right)^{2}=\left[\sum_{N=0}^{\infty} \left(\frac{t_2+t_{3}}{t_{1}}\right)^{N} \right]^2.\nonumber\\
\label{eq:hconverge}
\end{eqnarray}
Therefore for $\frac{t_2+t_{3}}{t_{1}}<1$, the normalization constant $Z_{\rm h}$ converges; namely the fully localized states appear. 

Next
we consider the diamond lattice with the $(111)$ surface in the half space $x+y+z \leq 0$.
The surface are located on $x+y+z=0$ plane, and the origin $\mathcal{O}$ is set to be $(0,0,0)$.
The wavefunction at $m \mathbf{d}_{1}+n\mathbf{d}_{2}+l\mathbf{d}_{3}$ in the 
A sublattice is denoted as $|A_{mnl}\rangle$, where $\mathbf{d}_i=\boldsymbol\tau_{i+1}-\boldsymbol\tau_{1}$ for $i=2,3,4$ (here the vectors $\boldsymbol{\tau}_{i}$ are the same as those in Sec III).
Then, the wavefunction of the fully localized state is expressed as
\begin{eqnarray}
|\Psi\rangle=\sum_{m,n,l}a_{mnl} |A_{mnl}\rangle,
\end{eqnarray}
 where $a_{mnl}$ is the amplitude for $|A_{mnl}\rangle$.
The Schr$\ddot{\mathrm{o}}$dinger equation leads to the relation between amplitudes as
\begin{eqnarray}
a_{mnl}=-\frac{t_2}{t_1}a_{m-1nl}-\frac{t_3}{t_1}a_{mn-1l}-\frac{t_4}{t_1}a_{mnl-1}.
\end{eqnarray}
Similarly to the case of the honeycomb lattice, the solution for the sequence is obtained by the shortest path problem in three dimensions.
Let $P_1$, $P_2$ and $P_3$ denote the probabilities for moving along $(1,0,0)$, $(0,1,0)$ and $(0,0,1)$ respectively. The probability $P_{mnl}$ of the shortest path problem from $\mathcal{O}$ to $(m,n,l)$ is given as
\begin{eqnarray}
P_{mnl}=\frac{(m+n+l)!}{m!n!l!}P_{1}^{m}P_{2}^{n}P_{3}^{l}.
\end{eqnarray}
By replacing the probabilities of the movement with the ratio of the hopping integrals, 
we have the amplitude $a_{mnl}$
as
\begin{eqnarray}
a_{mnl}=\frac{1}{Z_{\rm d}} \frac{(m+n+l)!}{m!n!l!}\left(-\frac{t_{2}}{t_1}\right)^{m} \left(-\frac{t_{3}}{t_1}\right)^{n}\left(-\frac{t_{4}}{t_1}\right)^{l},
\end{eqnarray}
where $Z_{\rm d}$ is the normalization constant for the wavefunction $|\Psi\rangle$, and this amplitude is nonzero only when $m$, $n$, $l$ are
all nonnegative.
The condition for existence of the fully localized state is that $Z_{\rm d}$ converges.
$Z_{\rm d}$ satisfies as the following relation
\begin{eqnarray}
Z_{\rm d}^2 &=&\sum_{m,n,k=0}^{\infty}P^2_{mnk}
=\sum_{N=0}^{\infty}\sum_{m,n,k=0}^{N}\delta_{m+n+k,N}P^2_{mnk}\nonumber\\
&\leq &\left(\sum_{N=0}^{\infty}\sum_{m,n,k=0}^{N}\delta_{m+n+k,N}|P_{mnk}|\right)^2
\nonumber \\ &=&\left[\sum_{N=0}^{\infty} \left(\frac{t_2+t_{3}+t_{4}}{t_{1}}\right)^{N} \right]^2.\label{eq:dconverge}
\end{eqnarray}
Therefore for $\frac{t_2+t_{3}+t_{4}}{t_{1}}<1$, the normalization constant $Z_{\rm d}$ converges, and the completely flat-band states appear.

Those convergence conditions for $Z_{\rm h}$ and $Z_{\rm d}$ agree with the conditions for existence of flat edge/surface states for the whole BZ.
However, 
we note that Eqs.~(\ref{eq:hconverge})(\ref{eq:dconverge}) give sufficient conditions for normalizability of the fully-localized wavefunction.
In Appendix~\ref{App:Zcalc}, we calculated $Z_{\rm h}$ and $Z_{\rm d}$ for $t_{1}=t$ and the others being unity, where $t$ is the hopping 
parameter of the anisotropy. The results agree with the conditions in Eqs.~(\ref{eq:hconverge})(\ref{eq:dconverge}).
Thus the fully localized states exist when there is a flat-band edge/surface-localized states for the whole BZ.

\section{Summary and Discussions}
Existence of flat-band boundary states is shown for tight-binding models
on the diamond lattice.
In the diamond-lattice model, we found that the verge of the distribution of the flat-band surface states is identical with the loop formed by the gap-closing points. 
Thus if the Fermi energy is zero, this loop corresponds to the Fermi surface.
We showed that the distribution of the bulk Fermi loop shrinks and then disappears in the BZ with increasing anisotropy of the hopping.
The surface flat bands cover the surface BZ completely when the anisotropy is sufficiently large.
These flat bands are understood topologically within the theory by Ryu and Hatsugai \cite{Ryu02}.
We found that the isotropic case is just at a 
topological phase transition of the bulk Fermi loop, and this transition is
driven by anisotropy of the hopping. 
Lastly, for strongly anisotropic cases we constructed a fully-localized wavefunction. 
The wavefunction is localized at a single site in
 the outermost edge/surface layer, while it expands inward with exponential decay.
From the Schr$\ddot{\mathrm{o}}$dinger equation, we calculated the wavefunction
of the spatially localized states both in the honeycomb- and diamond-lattice models,
and showed that they are normalizable for strongly anisotropic cases.

We note that our model is an idealized one; for example, only the nearest-neighbor hopping
is retained while other hopping is neglected. If other hopping is taken into account, 
the results for the flat-band boundary states will be modified. Nevertheless, as long as the
hopping to next-nearest neighbors and other sites are not so strong, the modification 
will be small. For example, in the graphene model, if we take into account next-nearest 
neighbor hopping, there will be a small dispersion to the otherwise flat (dispersionless)
edge states \cite{Sasaki06}. 
Though one might think that such an idealized model would be useless for real materials,
it is not the case. In fact, for the search of novel edge/surface states, idealized models discussed
in this paper work quite well. One can resort to 
first-principle calculations, only after candidate materials are identified. On the other hand, in order to search candidate materials, model calculations 
in this paper would be 
powerful in general and would give a hint to search for 
a new class of materials which have novel edge/surface states. 

For the flat-band edge states in graphene, a ferromagnetic magnetization 
is theoretically proposed when the Hubbard on-site interaction $U$ is included
\cite{Fujita96}. In this flat-band edge states, because within  
the flat-band states the kinetic energy is degenerate for all the multi-particle 
states, the Hubbard interaction favors the multi-particle states with spins 
all aligned parallel. Therefore, also in the flat-band surface states 
in the diamond-lattice model, magnetization is expected when the Hubbard 
on-site interaction is included. In reality, the flat-band states will be
dispersed by hoppings other than nearest-neighbor ones, and whether or not 
the magnetization depends on the relative size of the Hubbard $U$ 
versus the bandwidth for the (almost flat) surface states.

In previous works, the flat-band edge states in 2Ds have been studied, particularly in the context of graphene, while the flat-band surface 
states in 3Ds have not been explored in detail. 
In this paper we could explain the reason for existence of these boundary states in 2D and 3D in parallel and we also showed novel fully localized states both in 2D and 3D.
It would be interesting if there are materials realizing the flat-band surface states 
proposed in the present paper. 
While our proposed model is based on the diamond lattice, typical cubic semiconductors such as diamond
do not correspond to the proposed class of systems. In diamond there are four orbitals with sp$^3$ hybrid orbitals,
giving rise to a wide gap due to covalent bonding. In contrast, our tight-binding model has one s-like orbital per
site, giving rise to the gapless spectrum for isotropic systems. 
Moreover, to realize the completely flat band and the fully localized surface states,
strong anisotropy of hopping exceeding the factor of three is proposed, but it is 
too large to be accessible by external uniaxial pressure to isotropic systems. 
One can instead search for anisotropic systems from the outset. The corresponding structure would be 
a trigonal lattice structure, possibly in layered materials. Our tight-binding model (\ref{eq:TB}) 
applies also to this lattice structure. 
Although in 
layered materials, interlayer hopping is usually weaker than the intralayer hopping,
and it is the opposite to what we need for flat-band surface states.
Even in that case the flat-band surface states exist, even though the region for the flat-band
surface states is smaller than the isotropic case. 
Furthermore, a recent study has predicted a large anisotropy in the interlayer hopping in bilayer silicene by the ab-initio calculation \cite{Yao13}.
According to the prediction, the interlayer hopping is twice as large as the intralayer hopping without strain. 
Although the bilayer silicene does not have the chiral symmetry,
this example indicates possibility of existence of materials with large anisotropy in the hopping integrals.

In general, to obtain the localized boundary flat-band states, bipartite lattices are necessary. 
Bipartite-lattice models can be made in several ways: for example, by splitting the vertex of unipartite lattices.
In addition, those systems must have the anisotropy in the hopping along the direction normal to the surface.
Thus to find candidate materials for the flat-band surface states, one need to find bipartite systems with strong anisotropy, and when the bulk is gapped, it may have the complete flat band depending on the surface orientation. 

 
\begin{acknowledgments}
This work is partially supported by the Global Center of Excellence Program by MEXT, Japan through the ``Nanoscience and Quantum Physics'' Project of the Tokyo Institute of Technology, Grant-in-Aid from MEXT, Japan (No.~21000004), JSPS Research Fellowships for Young Scientists, and TIES, Tokyo Institute of 
Technology. We thank H. Katsura, M. Ezawa, and S. Nakosai for useful discussions, 
and P. Delplace for notifying us about Ref.~\onlinecite{Delplace11}.
\end{acknowledgments}

\appendix
\section{Calculations for normalization constants}\label{App:Zcalc}
In Appendix, we calculate the normalization constants, $Z_{\rm h}$ and $Z_{\rm d}$ in Section~\ref{sec:spatiallylocalizedstates}, for special cases where 
the hopping integrals perpendicular to the boundary is $t$ and the others are unity. 
\subsection{Honeycomb lattice}
For $t_{1}=t\equiv x^{-1}$ and $t_2=t_3=1$, 
$Z_{\rm h}$ is given as
\begin{eqnarray}
Z_{\rm h}^2&=&\sum_{m,n = 0}^{\infty} \left[ \frac{(m+n)!}{m!n!}(-x)^{m+n}\right]^2 \nonumber\\
&=&\sum_{N=0}^{\infty}\sum_{m= 0}^{N}\left[ \frac{N!}{m!(N-m)!}\right]^2 x^{2N}\nonumber\\
&=&\sum_{N=0}^{\infty}
\begin{pmatrix}
2N\\N
\end{pmatrix}
x^{2N}
= \frac{1}{\sqrt{1-4x^2}},
\end{eqnarray}
as long as $x=t^{-1}$ satisfies $|x|<\frac{1}{2}$.
Therefore for $|t|>2$ there are flat-band states which are localized 
on one site on the edge in the honeycomb lattice with zigzag edges.

\subsection{Diamond lattice}
For $t_{1}=t$ and $t_2=t_3=t_4=1$, $Z_{\rm d}$ is given as
\begin{eqnarray}
Z_{\rm d}^2&=&\sum_{m,n,k=0}^{\infty}\left(\frac{(n+m+k)!}{n!m!k!}\right)^2t^{-2(n+m+k)}\nonumber\\
&=&\sum_{N=0}^{\infty}\sum_{m,k=0}^{\infty}\left(\frac{N!}{(N-m-k)!m!k!}\right)^2t^{-2N}\nonumber\\
&=&\sum_{N=0}^{\infty}\oint_{C}\oint_{C}\frac{\mathrm{d}\xi\mathrm{d}\eta}{(2\pi i)^2 \xi\eta }(1+ \xi+ \eta )^N(1 +\xi^{-1}+ \eta^{-1} )^N
t^{-2N}\nonumber\\
&=&\oint_{C}\oint_{C}\frac{\mathrm{d}\xi\mathrm{d}\eta}{(2\pi i)^2 \xi\eta }\frac{1}
{1-(1+ \xi+ \eta )(1 +\xi^{-1}+ \eta^{-1} )Y^2},
\end{eqnarray}
where $t^{-1}=Y$ for notational brevity and we assume $0\leq Y \ll 1$. Later we can analytically continue the result with respect to $Y$ to discuss the condition for 
convergence of $Z_{\rm d}$. $\xi$, $\eta$ are complex numbers, and the path of the integrals is along the unit circle $C: |z|=1$.
In the above equation, by using $\xi=\mathrm{e}^{i\theta}$ and $\eta=\mathrm{e}^{i\phi}$ where $\theta,\phi \in \Re$, we have
\begin{eqnarray}
Z_{\rm d}^2 &=&\int_{0}^{2\pi}\frac{\mathrm{d}\theta\mathrm{d}\phi}{(2\pi)^2} 
 \frac{1}
{1-(1+ \mathrm{e}^{i\theta}+ \mathrm{e}^{i\phi})(1 +\mathrm{e}^{-i\theta}+ \mathrm{e}^{-i\phi})Y^2}\nonumber\\
&=&\int_{0}^{2\pi}\frac{\mathrm{d}\alpha\mathrm{d}\beta}{(2\pi)^2} 
\frac{1}
{1-(1+4\cos^2\beta +4\cos\alpha \cos\beta )Y^2}\nonumber\\
&=&\frac{1}{2\pi Y^2}\int_{0}^{2\pi}\frac{\mathrm{d}\beta }
{\sqrt{ (4\cos^2\beta-p^2)(4\cos^2\beta-q^2)}},
\end{eqnarray}
where $\alpha=\frac{\theta+\phi}{2}$, $\beta=\frac{\theta-\phi}{2}$, $p= 1+\frac{1}{Y}$, $q=\frac{1}{Y}-1$, and we use the relation $\int_{0}^{2\pi}\mathrm{d}\alpha\frac{1}{A+B\cos\alpha}=\frac{\mathrm{sgn}(A)}{\sqrt{A^2-B^2}}$ for $|A|>|B|$, and $A,B \in \Re$.
By changing the variable as $\tan\beta=v$, the above equation becomes
\begin{eqnarray}
Z_{\rm d}^2 &=&\frac{2}{\pi Y^2 pq}\int_{0}^{\infty}\mathrm{d}v \frac{1}
{\sqrt{ (v^2+b^2)(v^2+c^2)}},
\end{eqnarray}
where $b^2=1-\frac{4}{q^2}$ and $c^2=1-\frac{4}{p^2}$.
The integral is further transformed to the elliptic integral of the first kind,
\begin{eqnarray}
&&Z_{\rm d}^2 = C(Y)\int_{0}^{1}
 \frac{\mathrm{d}s}{\sqrt{(1-s^2)(1-k^2s^2)}}, \\
 &\ &\ \ \ \ C(Y)=\frac{2}{\pi Y^2 p q c }=\frac{2}{\pi\sqrt{(1-Y)^3(1+3Y)}}, \\
 &\ &\ \ \ \ k^2=\frac{16Y^3}{(1-Y)^3(3Y+1)},
\end{eqnarray}
where the variable is changed as $v=\frac{b s}{\sqrt{1-s^2}}$.
Therefore, for $0\leq Y<\frac{1}{3}$, $k^2$ satisfies $0\leq k^2<1$, and $Z_{\rm d}^2$ converges.
On the other hand, for $Y<0$, because $Z_{\rm d}^{2}$ is an even function of $Y$, $Z_{\rm d}^{2}$ converges for $-\frac{1}{3}<Y<0$.
Namely, the condition for the convergence of the normalization constant $Z_{\rm d}$ is $|t|<3$.


\begin{thebibliography}{99}
%

\bibitem{Lieb89}
 E. H. Lieb, Phys. Rev. Lett. \textbf{62}, 1201 (1989).
 \bibitem{Mielke91a}
 A. Mielke, J. Phys. A: Math. Gen. \textbf{24}, L73 (1991).
 \bibitem{Mielke91b}
A. Mielke, J. Phys. A\textbf{ 24}, 3311 (1991).

 \bibitem{Mielke91c}
A. Mielke, J. Phys. A\textbf{ 25}, 4335 (1992).


 \bibitem{Tasaki93}
A. Mielke and H. Tasaki, Commun. Math. Phys. \textbf{ 158}, 341 (1993).
\bibitem{Novoselov04}K. S. Novoselov {\it et al}., Science \textbf{306}, 666 (2004).
 \bibitem{Fujita96}
M. Fujita {\it et al.}, J. Phys. Soc. Jpn. \textbf{65}, 1920 (1996).
\bibitem{Ryu02}
S. Ryu, Y. Hatsugai, Phys. Rev. Lett. {\bf 89}, 077002 (2002).
\bibitem{Takagi08}
Y. Takagi and S. Okada, Surf. Sci. \textbf{602}, 2876 (2008).
\bibitem{Dietl08}
P. Dietl, F. Piechon, G. Montambaux, Phys. Rev. Lett. \textbf{100}, 236405 (2008).
\bibitem{Okada10}
M. Otani, M. Koshino, Y. Takagi, S. Okada,  Phys. Rev. B \textbf{81}, 161403 (2010).
\bibitem{Delplace11}
P. Delplace , D. Ullmo, G. Montambaux, Phys. Rev. B \textbf{84}, 195452 (2011).
\bibitem{Kohmoto07}
M. Kohmoto and Y. Hasegawa, Phys. Rev. B \textbf{76}, 205402 (2007).


\bibitem{Sasaki06}
K. Sasaki, S. Murakami, and R. Saito,
Appl. Phys. Lett. \textbf{88}, 113110 (2006).
\bibitem{Yao13}
Feng Liu {\it et al.}, Phys. Rev. Lett. \textbf{111}, 066804 (2013).











\end{thebibliography}

\end{document}